\documentclass[aps,prl,superscriptaddress,twocolumn,showpacs,amsmath,amssymb,floatfix]{revtex4}

\usepackage{graphicx}
\usepackage{dcolumn}
\usepackage{bm}

\newcommand{\kl}{\ensuremath{K_L}}
\newcommand{\ks}{\ensuremath{K^\ast}}
\newcommand{\pz}{\ensuremath{\pi^0}}
\newcommand{\pzd}{\ensuremath{\pi^0_D}}
\newcommand{\ee}{\ensuremath{e^+e^-}}
\newcommand{\mm}{\ensuremath{\mu^+\mu^-}}
\newcommand{\agg}{\ensuremath{\gamma\gamma}}
\newcommand{\eeg}{\ensuremath{\ee\gamma}}
\newcommand{\eegg}{\ensuremath{\ee\gamma\gamma}}
\newcommand{\eegpg}{\ensuremath{\ee\gamma(\gamma)}}
\newcommand{\tpd}{\ensuremath{\pz\pz\pzd}}
\newcommand{\gs}{\ensuremath{\gamma^\ast}}

\newcommand{\kmm}{\ensuremath{\kl\to\mm}}
\newcommand{\kmmg}{\ensuremath{\kl\to\mm\gamma}}
\newcommand{\keeg}{\ensuremath{\kl\to\eeg}}
\newcommand{\keegg}{\ensuremath{\kl\to\eegg}}
\newcommand{\keegpg}{\ensuremath{\kl\to\eegpg}}

\newcommand{\ktpd}{\ensuremath{\kl\to\tpd}}

\newcommand{\ktp}{\ensuremath{\kl\to\pz\pz\pz}}
\newcommand{\ket}{\ensuremath{\kl\to\pi^\pm e^\mp\nu_e}}

\newcommand{\kgg}{\ensuremath{\kl\to\agg}}

\newcommand{\pgg}{\ensuremath{\pz\to\agg}}
\newcommand{\peeg}{\ensuremath{\pz\to\eeg}}

\newcommand{\pdeegpg}{\ensuremath{\pzd\to\eegpg}}

\newcommand{\kgsg}{\ensuremath{\kl\gs\gamma}}
\newcommand{\kgsgs}{\ensuremath{\kl\gs\gs}}
\newcommand{\aks}{\ensuremath{\alpha_{\ks}}}
\newcommand{\caks}{\ensuremath{C\alpha_{\ks}}}
\newcommand{\adip}{\ensuremath{\alpha_{DIP}}}
\newcommand{\bdip}{\ensuremath{\beta_{DIP}}}
\newcommand{\vtd}{\ensuremath{V_{td}}}

\newcommand{\order}{\ensuremath{\cal{O}}}
\newcommand{\oas}{\order\ensuremath{(\alpha^2)}}
\newcommand{\oac}{\order\ensuremath{(\alpha^3)}}
\newcommand{\egcm}{\ensuremath{E_{\gamma 2}^{cm}}}
\newcommand{\mgg}{\ensuremath{m_{\agg}}}

\newcommand{\bms}{Bergstr\"{o}m, Mass\'{o}, and Singer}
\newcommand{\dip}{D'Ambrosio, Isidori, and Portol\'{e}s}

\newcommand{\UAz}{University of Arizona, Tucson, Arizona 85721}
\newcommand{\UCLA}{University of California at Los Angeles, Los Angeles,
                    California 90095} 
\newcommand{\UCSD}{University of California at San Diego, La Jolla,
                   California 92093} 
\newcommand{\EFI}{The Enrico Fermi Institute, The University of Chicago, 
                  Chicago, Illinois 60637}
\newcommand{\UB}{University of Colorado, Boulder, Colorado 80309}
\newcommand{\ELM}{Elmhurst College, Elmhurst, Illinois 60126}
\newcommand{\FNAL}{Fermi National Accelerator Laboratory, 
                   Batavia, Illinois 60510}
\newcommand{\Osaka}{Osaka University, Toyonaka, Osaka 560-0043 Japan} 
\newcommand{\Rice}{Rice University, Houston, Texas 77005}
\newcommand{\UVa}{The Department of Physics and Institute of Nuclear and 
                  Particle Physics, University of Virginia, 
                  Charlottesville, Virginia 22901}
\newcommand{\UW}{University of Wisconsin, Madison, Wisconsin 53706}

\begin{document}


\title{Measurements of the Decay \keeg}

\affiliation{\UAz}
\affiliation{\UCLA}
\affiliation{\UCSD}
\affiliation{\EFI}
\affiliation{\UB}
\affiliation{\ELM}
\affiliation{\FNAL}
\affiliation{\Osaka}
\affiliation{\Rice}
\affiliation{\UVa}
\affiliation{\UW}

\author{E.~Abouzaid}	  \affiliation{\EFI}
\author{M.~Arenton}       \affiliation{\UVa}
\author{A.R.~Barker}      \altaffiliation[Deceased.]{ } \affiliation{\UB}
\author{L.~Bellantoni}    \affiliation{\FNAL}
\author{E.~Blucher}       \affiliation{\EFI}
\author{G.J.~Bock}        \affiliation{\FNAL}
\author{E.~Cheu}          \affiliation{\UAz}
\author{R.~Coleman}       \affiliation{\FNAL}
\author{M.D.~Corcoran}    \affiliation{\Rice}
\author{B.~Cox}           \affiliation{\UVa}
\author{A.R.~Erwin}       \affiliation{\UW}
\author{A.~Glazov}        \affiliation{\EFI}
\author{A.~Golossanov}    \affiliation{\UVa}
\author{Y.B.~Hsiung}      \affiliation{\FNAL}
\author{D.A.~Jensen}      \affiliation{\FNAL}
\author{R.~Kessler}       \affiliation{\EFI}
\author{H.G.E.~Kobrak}    \affiliation{\UCSD}
\author{K.~Kotera}        \affiliation{\Osaka}
\author{J.~LaDue}         \affiliation{\UB}
\author{A.~Ledovskoy}     \affiliation{\UVa}
\author{P.L.~McBride}     \affiliation{\FNAL}

\author{E.~Monnier}
   \altaffiliation[Permanent address ]{C.P.P. Marseille/C.N.R.S., France}
   \affiliation{\EFI}

\author{H.~Nguyen}       \affiliation{\FNAL}
\author{R.~Niclasen}     \affiliation{\UB} 
\author{E.J.~Ramberg}    \affiliation{\FNAL}
\author{R.E.~Ray}        \affiliation{\FNAL}
\author{M.~Ronquest}	 \affiliation{\UVa}
\author{J.~Shields}      \affiliation{\UVa}
\author{W.~Slater}       \affiliation{\UCLA}
\author{D.~Smith}	 \affiliation{\UVa}
\author{N.~Solomey}      \affiliation{\EFI}
\author{E.C.~Swallow}    \affiliation{\EFI}\affiliation{\ELM}
\author{P.A.~Toale}      \affiliation{\UB}
\author{R.~Tschirhart}   \affiliation{\FNAL}
\author{Y.W.~Wah}        \affiliation{\EFI}
\author{J.~Wang}         \affiliation{\UAz}
\author{H.B.~White}      \affiliation{\FNAL}
\author{J.~Whitmore}     \affiliation{\FNAL}
\author{M.~J.~Wilking}
\altaffiliation[To whom correspondence should be addressed.]{ }
\affiliation{\UB}
\author{R.~Winston}      \affiliation{\EFI}
\author{E.T.~Worcester}  \affiliation{\EFI}
\author{T.~Yamanaka}     \affiliation{\Osaka}
\author{E.~D.~Zimmerman} \affiliation{\UB}

\begin{abstract}
The E799-II (KTeV) experiment at Fermilab has collected 83262 \keegpg\
events above a background of 79 events.  We measure a decay width,
normalized to the \ktpd\ (\pgg, \pgg, \pdeegpg) decay width, of
$\Gamma(\keegpg)/\Gamma(\ktpd) = \left(1.3302 \pm 0.0046_{stat} \pm
0.0102_{syst} \right) \times 10^{-3}$.  We also measure parameters of two
\kgsg\ form factor models.  In the \bms\ (BMS) parametrization, we find
$\caks = -0.517 \pm 0.030_{stat} \pm 0.022_{syst}$.  We separately fit for
the first parameter of the \dip\ (DIP) model and find $\adip = -1.729 \pm
0.043_{stat} \pm 0.028_{syst}$.
\end{abstract}


\maketitle

The rare decay \keeg\ offers a direct means for studying the dynamics of the
\kgsg\ vertex.  The form factor of this vertex is important for determining
the long-distance, two photon contribution to the \kmm\ decay width so that
the more interesting short-distance contributions can be determined.  These
contributions have been important for extracting a direct constraint on the
real part of the CKM matrix parameter \vtd~\cite{isidori,gorbahn}.  In
addition, this short-distance information provides one of the most stringent
constraints on flavor-changing neutral-current couplings of the Z boson,
which are beyond the standard model~\cite{isidori}.  In this letter we
present a measurement of the \kgsg\ form factor and \keeg\ branching ratio
using data from the 1997 run of the E799-II (KTeV) experiment at Fermilab.

Two parametrizations of the form factor are considered.  The first
model, proposed by \bms\ (BMS), describes the \kgsg\ vertex in terms
of two types of processes: a \kl\ transition into a \pz,$\eta$,$\eta'$
state that decays to two photons, and a vector meson dominance
contribution in which the \kl\ first decays into a \ks\ and a
photon followed by a strangeness changing vector-vector transition
into a $\rho$,$\omega$,$\phi$ state that decays into a virtual
photon~\cite{bms1}.  The relative size of these two contributions is
characterized by a constant parameter, $\caks$.  The BMS form factor
model, given in Eq. \ref{eq:fbms}, is a function of the Dalitz
variable $x$, which is defined as the squared ratio of the \ee\ mass
to the \kl\ mass.
\begin{widetext}
\begin{equation} \label{eq:fbms}
f_{BMS}(x)= \frac{1}{1-x\cdot\frac{M_K^2}{M_\rho^2}}
   +\frac{\caks}{1-x\cdot\frac{M_K^2}{M_{\ks}^2}}
   \cdot\left(\frac{4}{3}-\frac{1}{1-x\cdot\frac{M_K^2}{M_\rho^2}}
   -\frac{1}{9}\frac{1}{1-x\cdot\frac{M_K^2}{M_\omega^2}}
   -\frac{2}{9}\frac{1}{1-x\cdot\frac{M_K^2}{M_\phi^2}}\right)
\end{equation}

The form factor model proposed by \dip\ (DIP) is a slightly
more general model that applies to all \kgsgs vertices~\cite{dip1}.
\begin{equation} \label{eq:fdip}
f_{DIP}(x_1,x_2) = 1 + \adip\left(\frac{x_1}{x_1-\frac{M_{\rho}^2}{M_K^2}}
+\frac{x_2}{x_2-\frac{M_\rho^2}{M_K^2}}\right)
+\bdip\frac{x_1 x_2}{\left(x_1-\frac{M_\rho^2}{M_K^2}\right)
\left(x_2-\frac{M_\rho^2}{M_K^2}\right)}
\end{equation}
\end{widetext}
The variables $x_1$ and $x_2$ are the squared ratios of the masses of
the two virtual photons to the kaon mass.  The \keeg\ decay is only
sensitive to \adip\ since one of the photons emerging from the vertex
is real.

The E799 phase of the KTeV experiment focused 800 GeV protons from the
Tevatron at Fermilab onto a BeO target.  The resulting particles were
collimated into two parallel neutral beams.  The beams passed through
a 65 m vacuum decay region beginning 94 m downstream of the target.
Immediately following the decay region was a charged track
spectrometer.  It consisted of two upstream drift chambers, a dipole
magnet, and two downstream drift chambers.  The drift chambers
achieved a position resolution of about 100 $\mu$m, which corresponded
to a momentum resolution of $\sigma(P)/P=0.38\%~\oplus~0.016\%\cdot
P(\textrm{GeV}/c)$.  Both the decay pipe and charged spectrometer were
surrounded by lead-scintillator photon veto detectors.

A transition radiation detector (TRD) consisted of eight planes of
polypropylene felt, each followed by a multi-wire proportional chamber
(MWPC) containing an 80\%-20\% mixture of xenon and CO$_2$.  The TRD
provided a single track pion rejection of better than 200:1 at 90\%
electron efficiency~\cite{ggthesis}.

A CsI electromagnetic calorimeter was just downstream of the TRD.  The
calorimeter provided an energy resolution of
$\sigma(E)/E=0.45\%~\oplus~2\%/\sqrt{E(\textrm{GeV})}$.  Behind the
calorimeter was a muon system consisting of planes of lead, steel and
scintillator.  More detailed descriptions of the KTeV detector can be
found elsewhere~\cite{pee,pnn}.

A \keeg\ event was observed in the KTeV detector as two oppositely
charged tracks in the charged spectrometer that each pointed to a
cluster in the calorimeter, and a third calorimeter cluster
corresponding to the photon.  Each cluster was required to have an
energy greater than 2.75 GeV, and for electrons the ratio, $E/p$, of
cluster energy (measured by the calorimeter) to track momentum
(measured by the spectrometer) was required to lie between 0.925 and
1.075 to discriminate against muon and pion backgrounds.  Several cuts
were made to ensure that the decay vertex was inside the fiducial
region and to remove events near the edge of a detector.  A vertex
$\chi^2$ cut required that the decay particles originated from a
common vertex, and a $\chi^2$ cut on track matching at the center of
the analysis magnet ensured that the tracks were well-reconstructed.
The total kaon energy was required to be between 40 GeV and 200 GeV,
and less than 0.15 GeV of energy was allowed in each of the photon
veto counters.

Since all of the final state particles in this decay mode were
reconstructed, it was possible to impose two additional kinematic
constraints.  The first was a cut on the square of the component of
the reconstructed kaon momentum transverse to the original kaon
direction ($p_t^2$).  A cut on $p_t^2$ was placed at 500 (MeV/$c$)$^2$
to reduce events with extra particles not related to the decay and
events with missing decay particles.  The other kinematic constraint
was the requirement that the reconstructed kaon mass lie between 0.475
GeV/$c^2$ and 0.520 GeV$^2$.

\begin{figure}
\includegraphics[scale=0.45,clip=true]{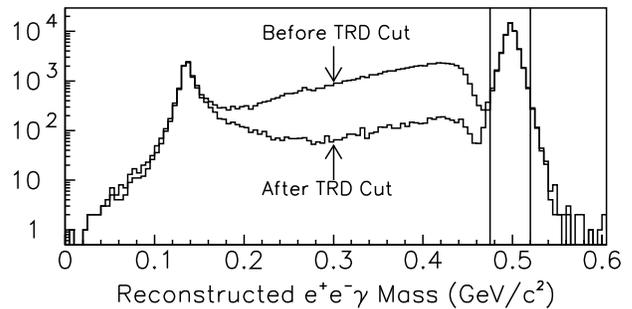}
\caption{\label{fig:dattrd} The reconstructed \eeg\ mass distribution
before and after the TRD cut shows a significant decrease in the
amount of background while retaining most of the signal.  The signal
region is indictated by the two vertical lines centered on the kaon
mass.}
\end{figure}

One of the main backgrounds was from \ket\ (K$_{e3}$) decays with either an
accidental or radiated photon.  In order for these events to mimic a signal,
the pion had to be misidentified as an electron. This occurred when most of
its energy was deposited in the calorimeter yielding an $E/p$ value close to
one.  The K$_{e3}$ background peaked below the signal in reconstructed \eeg\
mass, which is consistent with the kinematic limit of 0.477 GeV/$c^2$ for an
event with a pion misidentified as an electron.  This background was
significantly reduced by requiring a TRD pion probability of less than 5\%
for each track.  Fig. \ref{fig:dattrd} shows a plot of the \eeg\ invariant
mass before and after the TRD cut.

The other main background was due to \kgg\ events where one of the
photons converted into an \ee\ pair while exiting the vacuum region.
Since these photon conversions yielded \ee\ tracks that tended to be
close together upon reaching the charged spectrometer, requiring a
separation of 1.5 mm in the $x$ or $y$ view of the first drift chamber
of the charged spectometer effectively removed these events.  In a
Monte Carlo simulation of 10 times the expected number of \kgg\
conversion events, no events survived this cut.

To measure the branching ratio, the \keeg\ rate was normalized to the
rate of \ktpd\ (\ktp\ with two subsequent \pgg\ decays and one \peeg\
decay).  To reconstruct this mode, two tracks pointing to clusters in
the calorimeter and five additional photon clusters were required.  To
determine which four photons belonged to the two \pgg\ decays, the
decay vertex position was calculated for each pair of photons using
the separation distance and energies of the corresponding clusters,
and assuming that they came from a \pz.  A pairing $\chi^2$ was then
constructed based on the requirement that both \pz particles had the
same vertex position.  The photon pairing that yielded the smallest
$\chi^2$ was assumed to be correct.  The remaining photon was assumed
to come from the \peeg\ decay.  Photon mispairing occurred a small
fraction of the time, but this effect was well simulated in the Monte
Carlo.

The cuts used to isolate \ktpd\ events were kept as similar as
possible to those used for \keeg\ with a few exceptions.  The four
additional photons in the final state significantly improved the total
kaon mass resolution, so the kaon mass cuts were tightened to
0.485 GeV/$c^2$ and 0.510 GeV/$c^2$.  A loose cut was
made on the photon pairing $\chi^2$, and after matching the remaining
photon with the \ee\ pair, a cut was made on the \eeg\ mass to select
the region between 0.1275 GeV/$c^2$ and 0.1425 GeV/$c^2$.

The only significant background to the decay \ktpd\ was from \ktp\
events where one of the photons converted to an \ee\ pair.  This
background was removed with the same track separation requirement used
in the signal mode.

Monte Carlo simulations were used to determine the acceptances of the signal
and normalization decay modes.  Both modes used the same event generator as
the KTeV \keegg\ measurement~\cite{eegg}, which included \oas\ radiative
corrections in the calculation of the decay width~\cite{ms}.  Without
radiative corrections the measured branching ratio would have been shifted
by 1.7\%.

To implement \oas\ corrections, a cutoff of \mgg\ = 1 MeV/$c^2$ was
introduced to distinguish \keeg\ from the radiative decay, \keegg.
For events below the cutoff, the second photon was not generated.  The
value of the cutoff was chosen so that the energy of the softer photon
in \keegg\ events near the cutoff would be too low to be detected.
The detector acceptance for \keegpg\ events was found to be 3.4\%.
For \ktpd\ events, the acceptance was 0.26\%.

For the purposes of the branching ratio measurement, the decay \keeg\
can be defined as either all \keeg\ and \keegg\ events (the inclusive
definition), or by introducing a cutoff to distinguish \keeg\ from
\keegg\ (the exclusive definition).  We report our results using both
of these definitions.  For the exclusive result, \keeg\ is defined as
all events where the softer photon has an energy less than 5 MeV in
the kaon rest frame (\egcm\ $<$ 5 MeV).  This definition is chosen to
coincide with previous measurements of the \keegg\ branching
ratio~\cite{eegg}.

The same definitional ambiguity also exists for the normalization
mode.  Since previous measurements of \peeg\ have been
inclusive~\cite{peeg}, using the inclusive definition as the
normalization mode allows one to extract more easily the absolute
value of the \keeg\ branching ratio.

\begin{table}
\begin{center}
\begin{tabular}{|c|c|c|c|} \hline
Branching Ratio Uncertainties  & \% of BR & $\Delta$ \caks & $\Delta$ \adip \\
\hline \hline
Statistical Uncertainty        & 0.33\% & 0.030 & 0.038 \\ \hline
BR(\ktpd) Uncertainty          & 2.83\% & N/A   & N/A   \\ \hline
Absolute $\gamma$ Inefficiency & 0.43\% & N/A   & N/A   \\
Drift Chamber Inefficiency     & 0.37\% & 0.009 & 0.011 \\
Cut Variations                 & 0.33\% & 0.013 & 0.016 \\
Kaon Energy Spectrum           & 0.23\% & 0.011 & 0.014 \\
Trigger Inefficiency           & 0.21\% & N/A   & N/A   \\
Calorimeter Energy Resolution  & 0.14\% & 0.001 & 0.001 \\
Background Level               & 0.08\% & 0.000 & 0.000 \\
Detector Material              & 0.07\% & 0.008 & 0.009 \\
Drift Chamber Hit Resolution   & 0.04\% & 0.002 & 0.003 \\
\oac\ Radiative Corrections    & 0.03\% & 0.008 & 0.009 \\
Form Factor Dependence         & 0.03\% & N/A   & N/A   \\ \hline
Total Systematic Uncertainty   & 0.77\% & 0.022 & 0.028 \\ \hline
\end{tabular}
\end{center}
\caption{\label{tab:sys} Uncertainties for the \keeg\ branching ratio
and form factor parameter measurements.}
\end{table}

Since the normalization mode had four more photons than the signal mode, the
largest systematic uncertainty was the absolute photon inefficiency in the
calorimeter.  The three sources of photon detection bias considered were the
calorimeter geometry simulation, the simulated photon energy spectrum, and
the electromagnetic shower containment.  To measure the effect of the
simulated calorimeter geometry, the outer edge of the calorimeter was moved
by 0.5 mm in the \ktpd\ Monte Carlo events only.  This resulted in a 0.226\%
variation in the branching ratio.  The photon energy simulation was tested
by shifting the Monte Carlo photon energies by 10 MeV, which caused the
branching ratio to vary by 0.219\%.  These variations were chosen based on
detector survey data and information from other decay modes.  Finally, an
upper bound on the effect of imperfect electromagnetic shower containment in
the calorimeter was found by studying the low-end tail of the $E/p$
distribution for electrons.  Both \ket\ events and \ktp\ events were used to
find best-fit shapes for the low-end $E/p$ tail.  Switching between these
two shapes in the Monte Carlo caused a variation of 0.288\% in the branching
ratio.  These three effects, added in quadrature, resulted in a 0.43\%
systematic uncertainty.

The inefficiency of the drift chambers was also a source of systematic
uncertainty.  Two-dimensional inefficiency maps of each drift chamber
were measured using \ket\ decays.  These inefficiency maps were then
adjusted by a constant factor to give the best agreement between data
and Monte Carlo. One sigma variations about the best-fit value for
this factor yielded a 0.37\% systematic error on the branching ratio.

An overall systematic uncertainty due to disagreements between the
data and Monte Carlo was measured by studying the effect of varying
the analysis cuts.  The cut values on many important quantites were
studied, such as the vertex position, $E/p$, energy in the photon
vetos, reconstructed mass, pion probability, minimum photon energy,
and total energy.  These variations resulted in a 0.33\% variation in
the branching ratio.

There was also a systematic uncertainty due to the simulation of the
kaon energy spectrum.  The ratio of the total kaon energy
distributions between data and Monte Carlo exhibited a slope in both
the signal and normalization modes.  The slope was corrected by
reweighting the Monte Carlo events based on total kaon energy.  The
reweighted Monte Carlo events were used for the central value
measurement, and the variation in the branching ratio due to
reweighting, 0.23\%, was treated as a systematic uncertainty.

Several other smaller systematic effects have been evaluated as well.
The full list of systematic uncertainties is given in Table
\ref{tab:sys}.

We have observed 83,262 \keegpg\ events over a background of 79
events.  Using 4,924,801 \ktpd\ events to normalize the \keegpg\ rate,
the inclusive (\keeg\ + \keegg) and exclusive (\egcm $<$ 5 MeV) ratios
of the two \kl\ decay widths have been measured to be
\begin{eqnarray*}
\frac{\Gamma(\eeg)_{inc}}{\Gamma(\tpd)_{inc}} & = & (1.3302 \pm 0.0046
\pm 0.0102) \times 10^{-3} \\
\frac{\Gamma(\eeg)_{exc}}{\Gamma(\tpd)_{inc}} & = & (1.2521 \pm 0.0044
\pm 0.0097) \times 10^{-3} \\
\frac{\Gamma(\eeg)_{exc}}{\Gamma(\tpd)_{exc}} & = & (1.2798 \pm 0.0045
\pm 0.0099) \times 10^{-3}
\end{eqnarray*}
Using the known values of BR(\ktp), BR(\pgg) and
BR(\peeg)$_{inc}$~\cite{PDBook}, the absolute \keeg\ branching ratio is
\begin{eqnarray*}
BR_{inc} &=& (9.128 \pm 0.032 \pm 0.070 \pm 0.252) \times 10^{-6} \\
BR_{exc} &=& (8.591 \pm 0.030 \pm 0.066 \pm 0.238) \times 10^{-6}
\end{eqnarray*}
The third error listed is an external systematic error due
to the uncertainty in the \ktpd\ branching ratio.  These results are
in good agreement with all previous measurements with a factor of 3
improvement in the uncertainty on the ratio~\cite{bnle845,na31,na48}.

\begin{figure}
\includegraphics[scale=0.45,clip=true]{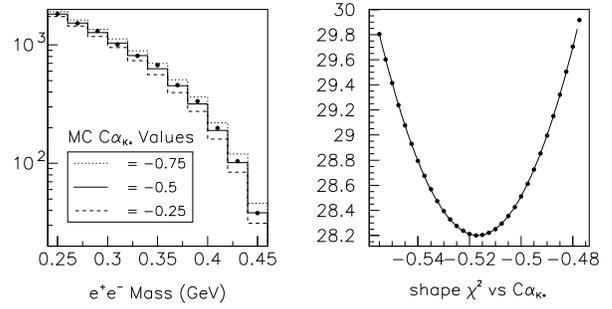}
\caption{\label{fig:meeform} The variation in the shape of the \ee\ mass
distribution was used to determine the values of the form factor parameters.
Three comparisons between data (dots) and Monte Carlo samples (histograms)
with different values of \caks\ are shown as well as the final fit to the
shape $\chi^2$ values}
\end{figure}

Since both form factor models being considered are functions of the
Dalitz variable $x$, the shape of the \ee\ mass distribution is very
sensitive to the form factor parameters.  Thus, each form factor
parameter was extracted by comparing the shape of the \ee\ mass
spectrum in data to several Monte Carlo samples with differing form
factor parameter values.  A few such comparisons are shown in
Fig. \ref{fig:meeform}.  For each comparison, a bin-by-bin
shape-$\chi^2$ value was calculated.  A quadratic fit to these
$\chi^2$ values was used to determine each form factor parameter.
This fitting procedure was performed separately for the two form
factor parameters since the two models depend on $x$ differently.

The uncertainties associated with the form factor measurement are
shown in Table \ref{tab:sys}.  Unlike the branching ratio
uncertainties, the form factor uncertainties are reported as a
variation in the measurement, not as a percentage of the final result,
since a measurement of zero has no special significance.  The method
for determining these uncertainties was the same as for the branching
ratio.

\begin{figure}
\includegraphics[scale=0.45,clip=true]{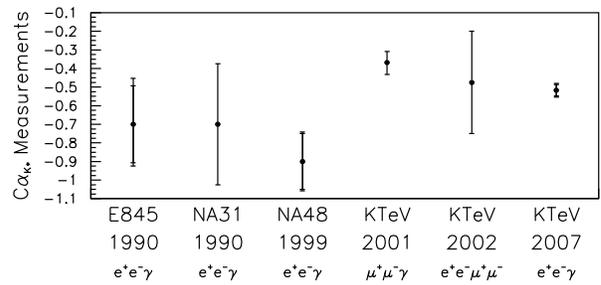}
\caption{\label{fig:akscomp} The graph shows a comparison of our measurement
of \caks\ with previous measurements.  The outer error bars represent the
total uncertainty and the inner error bars are the statistical uncertainty
only (where applicable)~\cite{bnle845,na31,na48,mmg,eemm}}
\end{figure}

The final fit for the form factor parameters \caks\ and \adip\ yield
\begin{eqnarray*}
\caks & = & -0.517 \pm 0.030 \pm 0.022 \\
\adip & = & -1.729 \pm 0.043 \pm 0.028
\end{eqnarray*}
The first uncertainty is from the $\chi^2$ minimization fit and the
second is the total systematic uncertainty.

Previous measurements of the BMS form factor have been reported in
terms of \aks, where the constant parameter $C$ has been divided out.
Since the expression for $C$ involves several experimental quantities
that have not been treated consistently in the past, we report \caks\
to avoid confusion.  Using previous measurements of \aks\ and their
corresponding values of $C$, a comparison with our measured value of
\caks\ is shown in Fig. \ref{fig:akscomp}.  In recent years, the
proper value for \caks\ has been unclear due to a 3.1$\sigma$
difference between the two previous best measurements.  Our
measurement lies between these values:  2.0$\sigma$ below the
KTeV \kmmg\ result and 2.4$\sigma$ above the NA48 \keeg\ result.

\bibliography{eeg_prl}
\end{document}